# A Complete End-To-End Open Source Toolchain for the Versatile Video Coding (VVC) Standard

Adam Wieckowski[*], Christian Lehmann[*], Benjamin Bross[*], Detlev Marpe[*],
Thibaud Biatek[+], Mickael Raulet[+], Jean Le Feuvre[$]

[*]Video Communication and Applications Department, Fraunhofer HHI, Berlin, Germany
[+]ATEME, Vélizy-Villacoublay, France
[$]LTCI, Telecom Paris, Institut Polytechnique de Paris, France
{firstname.lastname}@hhi.fraunhofer.de, {t.biatek,m.raulet}@ateme.com, jean.lefeuvre@telecom-paris.fr

## ABSTRACT

Versatile Video Coding (VVC) is the most recent international video coding standard jointly developed by ITU-T and ISO/IEC, which has been finalized in July 2020. VVC allows for significant bit-rate reductions around 50% for the same subjective video quality compared to its predecessor, High Efficiency Video Coding (HEVC). One year after finalization, VVC support in devices and chipsets is still under development, which is aligned with the typical development cycles of new video coding standards. This paper presents open-source software packages that allow building a complete VVC end-to-end toolchain already one year after its finalization. This includes the Fraunhofer HHI VVenC library for fast and efficient VVC encoding as well as HHI's VVdeC library for live decoding. An experimental integration of VVC in the GPAC software tools and FFmpeg media framework allows packaging VVC bitstreams, e.g. encoded with VVenC, in MP4 file format and using DASH for content creation and streaming. The integration of VVdeC allows playback on the receiver. Given these packages, step-by-step tutorials are provided for two possible application scenarios: VVC file encoding plus playback and adaptive streaming with DASH.

## CCS CONCEPTS

• Computing methodologies~Computer graphics~Image compression • Software and its engineering~Software creation and management~Collaboration in software development~Open source model

## 1 Introduction

In July 2021, the Versatile Video Coding (VVC) standard has been finalized by the Joint Video Experts Team (JVET) of ITU-T VCEG and ISO/IEC MPEG [1][2]. VVC was designed with two main objectives: significant bit-rate reduction over its predecessor High Efficiency Video Coding (HEVC) for the same perceived video quality and versatility to facilitate coding and transport for a wide range of applications and content types. This ranges from adaptive streaming of screen content to 360-degree video for virtual reality. The coding efficiency improvement of VVC over HEVC has been verified by independent subjective testing with naive viewers resulting in over 40% bit-rate reduction for High-Definition (HD) and Ultra-HD (UHD) video [3][4] as well as over 50% bit-rate reduction for 360-degree video [4] on average. For standard development and verification, the VVC test model (VTM) reference software has been used.

VTM is well suited to evaluate the compression potential of VVC but not optimized for runtime and lacks important real-world features on the encoder side such as rate control, subjective optimizations, parallelization as well different efficiency / runtime trade-offs. Furthermore, VVC specifies the bitstream of coded video as well as the decoding processes. To use a video coding standard in real world application scenarios, its integration into transport systems layer such as MPEG-2 transport stream (MPEG2-TS) [5], ISO base media file format (ISOBMFF a.k.a. MP4 file format) [6] tracks is key. In addition, VVC encapsulated in MP4 file format can be used in dynamic adaptive streaming over HTTP (DASH) [7].

Real-world VVC implementations in chipsets and devices together with mp4 file format support for file-based playback and streaming typically emerge 2-3 years after finalization of a standard. However, already today open-source software packages allow to use VVC with Fraunhofer HHI's encoder VVenC and decoder VVdeC. VVC bitstreams can be encapsulated in mp4 file format and packetized into DASH segments using GPAC while extraction and playback is feasible with VVdeC integrated into FFmpeg and GPAC. This paper describes these software packages in Section 2. Section 3 demonstrates how to use them for file playback and DASH streaming applications. Finally, Section 4 concludes this paper and gives an outlook of future work.

## 2 VVC Toolchain Software Packages

### 2.1 VVenC

The Fraunhofer Versatile Video Encoder, VVenC, is an optimized software encoder implementation of the VVC standard. Its source code is freely available on GitHub under a 3-clause BSD copyright license for both commercial and non-commercial use [8]. After the initial release in September 2020, which is only two months after finalization of VVC, version 1.0.0 has been released in May 2021, for which the results are presented.

VVenC has initially been derived from the VTM reference software. Its main purpose is to maintain the high coding efficiency of VTM while running significantly faster and providing real-world application features. While VVenC itself is implemented in C++, the software package includes a C library interface as well as two sample applications that can be used as standalone command line encoders – a simple application designed for ease-of-use as well as an expert app resembling the VTM interface. VTM as well as VVenC convert a raw YUV video input into a VVC compliant bitstream.

*2.1.1. Main features.* Compared to the VTM software the encoder is much more user friendly. This includes five pre-configured **presets** allowing different encoding speed / compression efficiency trade-offs: *faster*, *fast*, *medium*, *slow* and *slower*. While the slower preset delivers full VTM compression performance at less than half of its runtime, other presets provide faster and much faster encoding at the cost of bitrate increase. Additional speedup can be obtained using **multi-threading** with minimal compression efficiency degradation. A combination of coding tree unit line and frame parallel processing provides very good thread number to runtime reduction scaling [9]. Single-pass and two-pass frame-based **rate control** modes have been implemented, allowing very efficient encoding at given bit rates. Both have also been optimized for multi-threaded operation, including frame-level parallelism without performance impact. VVenC further incorporates **subjective optimizations** by using a weighted extended PSNR distortion metric (XPSNR). This overcomes the limited correlation of commonly used sum-of-squares- (SSD) or sum-of-absolutes (SAD) distortion calculations with perceived distortions using a simplified model of the human visual system to calculate weights. Furthermore, XPSNR can be calculated block-wise and therefore incorporated into block-level rate-distortion decisions. It is implemented by the means of local quantization parameter adaptations (QPA) and designed to improve the rate control accuracy and performance as well [10]. As of the current version, VVenC is compliant to the **Main 10 profile** (4:2:0 chroma subsampling and up to 10 bit per sample) and optimized for standard dynamic range (SDR) as well as for **high dynamic range (HDR)** content with wide color gamut. Currently, only one slice and tile per picture is supported, without sub-pictures. VVenC already anticipates versatile application scenarios by incorporating all Main 10 profile **screen content coding** tools and providing a constrained encoding mode optimized for use in open-GOP **adaptive streaming with resolution change** [11].

*2.1.2. Results.* The multi-threaded runtime and PSNR-based Bjøntegard Delta (BD) rate compression performance (negative numbers mean bit-rate reduction) for VVenC 1.0.0 relative to the HEVC reference software (HM-16.22) is shown in Figure 1. The results are based on encoding of the HD and UHD sequences from the JVET common test conditions [12] test-set under JVET random access conditions. Additionally, the figure contains results for VTM 12.0 as well as for two alternative freely available encoders, for HEVC (x265 3.4) and AV1 (aomenc 3.0). The results show that VVenC can achieve VTM coding efficiency at much lower runtimes when run multi-threaded. Compared to aomenc, it provides a steady 20% bit-rate reduction at similar runtimes. The three faster presets of VVenC can match the three slowest x265 presets regarding encoding time at very high bit-rate savings of around 50%. However, there are no VVenC presets available yet to match the encoding speed in faster x265 presets.

In recent JVET VVC verification tests, initial versions of VVenC were subjectively tested together with VTM and HM for UHD (VVenC v0.1) and HD (VVenC v0.3). These tests, which have been conducted in a controlled laboratory environment with naive viewers, concluded that VVenC in medium preset with subjective optimizations visually outperformed VTM.

### 2.2 VVdeC

The Fraunhofer Versatile Video Decoder, VVdeC, is an optimized VVC software decoder implementation which is freely available on GitHub under a 3-clause BSD copyright license [13]. Analogue to VVenC, the license covers both commercial and non-commercial use. The latest release, v1.1.1, was published in May 2021. It is compliant with the VVC **Main 10 profile** and correctly decodes the current VVC conformance testing set [14].

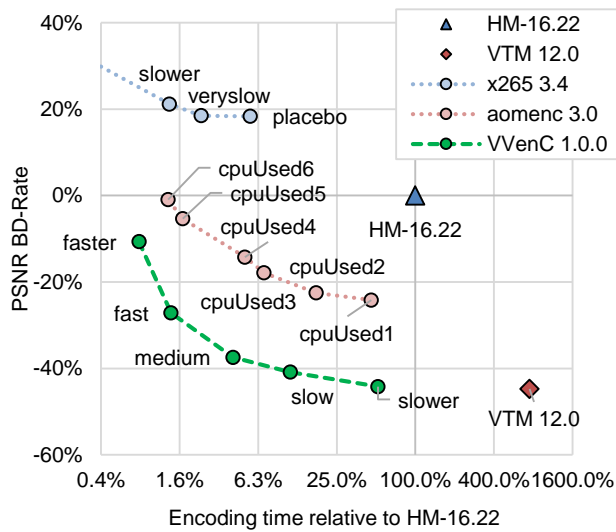

**Figure 1: Runtime and BD rate VVenC 1.0.0, x265 3.4, aomenc 3.0 and VTM 12.0 compared to HM 16.22. All codecs but VTM and HM are run multithreaded with 8 threads.**

The decoder software has been derived from VTM as well, with subsequent optimizations and parallelization. Being based on VTM, it has performance limitations beyond the complexity of the VVC standard itself. The decoder allows playback of HD video at 60 frames per second (fps) on most modern computers using only 2 or 3 threads, and **60 fps UHD playback** on more powerful modern workstations with sufficient processing cores to fully exploit the multi-threading potential [15].

The VVdeC package contains a simple and easy-to-use C library interface and a standalone decoder application capable of decoding elementary VVC bitstreams into raw YUV video data. The framework integrations described in the following sections allow actual video playback.

## 2.3 FFmpeg VVC integration with VVdeC

FFmpeg is a well-known multimedia framework which provides a set of libraries to record, convert and stream any type of audio and video data. In order to support VVC in FFmpeg, a parser and a Code Bitstreams Type (CBT) were implemented in the development branch (currently under review). This first step enables support for VVC elementary streams but does not provide a system level support with standard delivery mechanisms such as MPEG2-TS or DASH. To further extend the VVC capabilities in FFmpeg, the support for VVdeC and MPEG2-TS as well as ISOBMFF/DASH binds have been implemented by ATEME in a public FFmpeg fork [16]. The support for VVC video stream type has been added to MPEG2-TS demuxer in the libavformat. The ISOBMFF demuxer has been upgraded to support VVC tag 'vvc1' and 'vvcC' MOV table entry. Finally, the VVC MP4-to-AnnexB filter has been implemented in libavcodec for proper demuxing and decoding.

The provided implementation enables FFmpeg to natively support VVC services when delivered in MPEG2-TS or DASH. In addition, it provides VVC support through libavcodec when integrated in other projects, such as GPAC.

Within this fork, ffplay can be used to playback DASH manifest, or UDP MPEG2-TS streams, by using:

```
ffplay -i URL/manifest.mpd
ffplay -f mpegts -i udp://host:port
```

## 2.4 VVC support in GPAC

The GPAC multimedia framework provides a set of tools to package, stream and playback multimedia content. It is well-known for its MP4Box tool, an mp4 file packager and HTTP streaming (HLS) segmenter, but GPAC also enables end users to build custom multimedia processing pipelines through its filter-based architecture [17].

The systems aspect of VVC have been implemented in the master branch of GPAC and are part of the nightly builds of the project. This covers the most common application use cases:

- MPEG-2 broadcasting: multiplexing and demultiplexing an MPEG2-TS program with VVC content
- MP4 file packaging: importing and dumping VVC bitstreams
- MPEG-DASH and HLS content packaging

GPAC supports all codecs integrated in FFmpeg's libavcodec library, including VVdeC as included in [16]. This is detected at configure time and currently requires a custom build of GPAC, VVdeC and libavcodec, as previously explained. The GPAC player embeds the customized libavcodec which itself embeds the VVdeC core decoding library.

For VVC encoding and until integration of VVenC is finalized in libavcodec, GPAC can consume the VVenC bitstream output through files and pipes.

## 3 Application scenarios

In this section, we give some examples of using VVC in common case scenarios. The described end-to-end chain addresses both MPEG2-TS and IP based delivery mechanisms.

When used in a traditional digital video broadcast terrestrial/satellite environment (DVB-T2/S2), VVC bitstreams encapsulated in MPEG2-TS can be efficiently delivered and played with the proposed framework. The FFmpeg player has been used for that purpose during an in-field VVC broadcasting trial [18] showing interoperability with legacy delivery infrastructure. Recent full-IP broadcasting technologies such as ATSC-3.0 or the currently under-standardization TV3.0 in Brazil have also been addressed. The VVC encapsulation and carriage with ISOBMFF opens the door for such modern applications where the proposed solution can be used in an end-to-end manner. VVenC can be used to prepare and encode the video sequence (Figure 2), while GPAC can package and multicast the services using DASH/ROUTE, in an ATSC-3.0 fashion. At the receiver side, the GPAC client integrated with libavcodec/VVdeC can be used to deliver TV services to the viewer (Figure 2). The proposed solution is currently part of a response to the SBTVD TV3.0 [19] call for technology, utilizing presented components for playback.

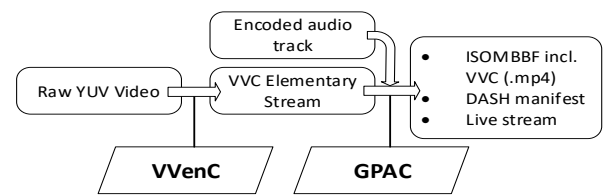

a) Encoding workflow and used components.

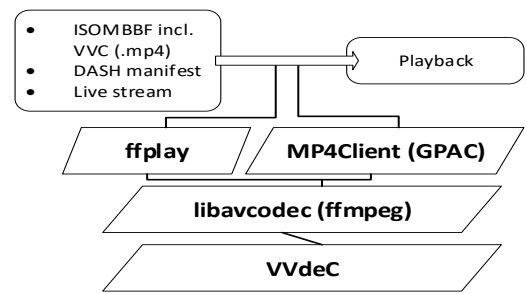

b) Decoding workflow and components.

**Figure 2: Encoding and decoding workflows utilizing the described components.**

In the following subsections two encoding and distribution scenarios using VVC are demonstrated step-by-step.

### 3.1 VVC encoding and packaging for playback

VVenC requires a YUV input file, either existing or decoded from another video file. Starting from a source video file, the following steps are required to create a packaged VVC transcode of the non-VVC input *input.mp4*:

- (Optional) decode an existing video to 8-bit YUV *iterm.yuv*
  ```
  gpac -i input.mp4 -o iterm.yuv
  ```
- Encode the raw YUV data to VVC elementary stream *es.vvc*
  ```
  vvencapp --preset medium -i iterm.yuv -s 1920x1080 [OPTS] -o es.vvc
  ```
- Possible packaging options with GPAC
  - Package the ES as MP4 to *VVC_demo.mp4*
    ```
    MP4Box -add es.vvc -new VVC_demo.mp4
    ```
  - Mux raw VVC and AAC streams to MP4
    ```
    MP4Box -add es.vvc -add audio.aac -new VVC_demo.mp4
    ```
  - Create an MPEG-2 TS live-stream at 10mbps (VVenC bitstreams might exceed some rate constraints)
    ```
    gpac -i es.vvc -i audio.aac -o udp://127.0.0.1:1234/:ext=ts:rate=10m:realtime
    ```

In the presented examples the VVC elementary stream (ES) file extension does not matter, because GPAC including MP4Box always probe the input data.

The following commands enable MP4 playback and high-level inspection including VVC bitstream:
```
MP4Client -gui URL
gpac -i URL inspect:deep:anayze=bs
```

### 3.2 VVC encoding and packaging for DASH

HTTP streaming works by segmenting the input media into short segments, each starting with an IDR frame. Some configurations will expect a roughly constant duration of each segment, which may require enforcing fixed frame structure. VVenC can be forced to create appropriate encoding for DASH using the parameters *--refreshtype idr*.

GPAC accepts both raw bitstreams or packaged files, such as MP4, as input to its DASH preparation process. The creation of DASH sessions in different profiles can be acchieved using:

- Using the "onDemand" profile:
  ```
  MP4Box -dash 2000 -profile onDemand es.vvc -out dash/vod.mpd
  ```
- Using the "live" profile:
  ```
  MP4Box -dash 2000 -profile live es.vvc -out dash/vod.mpd:dual
  ```
- Using segment timeline for variable duration segments:
  ```
  MP4Box -dash 2000 -profile live es.vvc -out dash/vod.mpd:stl
  ```

The following illustrates creating live DASH and HLS sessions made available through GPAC HTTP server:
```
gpac flist:srcs=VVC_AAC.mp4:flop=-1 reframer:rt=on @ -o http://127.0.0.1:8080/live.mpd:gpac:dual:segdur=2:cdur=0.1:asto=1.9:llhls=br:dmode=dynamic:rdirs=dash_live
```

More information regarding the complex subject of HTTP Streaming is available on GPAC's wiki, including *HowTos*.

## 4 Conclusion and Future Work

This paper teaches how to use publicly available open-source software to set up a complete VVC toolchain for file playback and adaptive streaming application scenarios.

Beyond these more traditional use cases, future work will focus on novel applications like adaptive streaming utilizing resolution change or high-resolution streaming with tiles for VR applications.